\documentclass[twocolumn]{aastex701}
\usepackage{amsmath}
\newcommand{\rg}{r_{\rm g}}
\newcommand{\Mdot}{\dot{M}}
\newcommand{\mdot}{\dot{m}}
\newcommand{\Edd}{_{\rm Edd}}
\newcommand{\trap}{_{\rm trap}}

\begin{document}

\title{Constraining Accretion Regimes of Little Red Dots from Their Variability}

\author[orcid=0000-0002-4064-0446,gname=Indu K. ,sname='Dihingia']{Indu  K. Dihingia}
\affiliation{Institute of Fundamental Physics and Quantum Technology,\& School of Physical Science and Technology, Ningbo University, Ningbo, Zhejiang 315211, People’s Republic of China}
\affiliation{Zhejiang Key Laboratory of Extreme Universe, Ningbo, Zhejiang 315211, People’s Republic of China}
\affiliation{Tsung-Dao Lee Institute, Shanghai Jiao Tong University, 1 Lisuo Road, Shanghai, 201210, People’s Republic of China}
\email[show]{ikd4638@gmail.com, indukalpa@nbu.edu.cn}

\begin{abstract}
The ``little red dots'' (LRDs) are photometrically and spectroscopically non-variable on rest-frame baselines of months, with a $5.9\sigma$ deficit relative to ordinary active galactic nuclei (AGNs). If we consider LRDs to be accreting black holes, this non-varying could be a clue about the nature of the accretion. Here, we present a general semi-analytic model of variability from super-Eddington (slim) accretion disks. We consider three mechanisms that play crucial roles in the observed variability, viz., advective attenuation, photospheric averaging, and photon diffusion filter. With this, above $\mdot \sim$ a few, the fractional rms collapses steeply with $\mdot$. Comparing with JWST monitoring constraints, we find that the non-detection of LRD variability requires $\mdot \gtrsim 30$ for $M_{\rm BH}\sim10^{7}\,M_\odot$ on a 0.5\,yr rest-frame baseline; a Monte Carlo forward model of the TWINKLE non-detections of 18 LRDs sharpens this to $\mdot\gtrsim60$ at 95\% confidence. On current baselines, these non-detections remain formally degenerate  with a frozen sub-Eddington disk ($\mdot\lesssim0.1$). Longer monitoring  will separate the two regimes as the frozen-disk variability grows with baseline, while the super-Eddington collapse does not. The same framework predicts strongly suppressed X-ray variability, testable with future X-ray observatories.

\end{abstract}



\section{Introduction} \label{sec:intro}

Early JWST imaging survey revealed an unexpected population of compact, extremely red astrophysical sources at redshift $\sim4$--$8$ \citep{labbe23}. Further, NIRSpec spectroscopy showed that many host broad H$\alpha$ lines, which are indicative of accreting black holes \citep{harikane23, kocevski23, onoue23}. The name coined as ``little red dots'' (LRDs; \citealt{matthee24}), however, their nature remains debated since their discovery \citep{labbe25, greene24, kocevski25, akins25}. If they host accreting black holes, several of their properties are implausible: weaker X-ray \citep{ananna24, kokubo25}, no dust torus \citep{casey24}, and unexpectedly no detectable variability. Multi-epoch NIRCam imaging of $\sim$300 LRDs suggested the presence of variability in only a few percent of sources \citep{zhang25}. Subsequently, \citet{kokubo25} and \citet{tee25} found no variability in small spectroscopic samples and in the rest-frame UV, respectively. Very recently, the JWST/TWINKLE slitless spectroscopic monitoring program detected zero variable sources among 18 LRDs at redshift $\sim3.9$--$6.8$ over rest-frame baselines of 140--220 days; if LRDs varied like reverberation-mapped AGNs, observing zero variable sources amounts to a $5.9\sigma$ deficit \citep{liu26}.

Super-Eddington accretion is a leading interpretation for LRDs \citep{inayoshi25a, inayoshi25b, madau26}. Radiation-MHD simulations of super-Eddington accretion disks produce remarkably steady light curves \citep{secunda26}.  \citet{inayoshi25c} also argue that photon trapping suppresses the X-ray and UV/optical variability from high-redshift AGNs, which is consistent with the TWINKLE comparison to simulated light curves \citep{liu26, secunda26}. What is missing is a general, analytic framework to answer the quantitative questions that the observations now pose, viz., at which $\mdot$ does variability disappear, how does it depend on black hole mass, wavelength, and baseline, and most importantly, what would falsify the Super-Eddington picture for LRDs? The current existing results are tied to specific simulations at specific parameters, so they can only hint but are unable to answer these questions concretely and quantitatively.

In this work we provide such a semi-analytical framework by using the radial structure of Super-Eddington (slim) disks with winds. In this model, we identify three damping mechanisms that act on accretion-rate fluctuations and assemble them into a single variability estimator calibrated using ordinary quasars (Section~\ref{sec:math}). The resulting ``maximum observable variability'' turns the LRD non-detections into a limit on $\mdot$, predicts mass segregation of the variable minority (Section~\ref{sec:results}), and this framework can be extended naturally to the X-ray band (Section~\ref{sec:disc}).

\section{Mathematical Background} \label{sec:math}

We adopt the standard supercritical accretion disk following \citep{ss73, abramowicz88, watarai99}, where matter with an accretion rate $\Mdot_0 = \mdot\,\Mdot\Edd$ (with $\Mdot\Edd \equiv L\Edd/0.1c^{2}$) is supplied at a large radius. Additionally, we considered a wind expelled inside the spherization radius $r_{\rm sph} = \mdot R_S$ (with $R_S = 2GM_{\rm BH}/c^{2}$; the value of $r_{\rm sph}$ may reach $\sim$$13\,\mdot\,\rg$ ($\rg=GM_{\rm BH}/c^{2}$) considering different conventions), depleting the disk at a rate $\Mdot(r) = \Mdot_0\,(r/r_{\rm sph})^{s}$ with $s=1$ fiducial. Throughout the work, $G, M_{\rm BH}$ and $c$ correspond to universal gravitational constant, mass of the black hole, and speed of the light, respectively. This value is chosen because, the local super-Eddington factor scales as $F_{\rm visc}/F_{\rm Edd}\propto\Mdot(r)/r\propto r^{s-1}$ ($r$ being radial distance from the black hole), and therefore $s=1$ is the self-consistent profile in which the disk remains at the marginally driven state throughout the trapping region \citep{ss73, poutanen07}. The higher values ($s>1$) would drop the inner disk below Eddington and switch off the radiation driving itself; similarly, smaller values ($s<1$) are physically allowed.
We introduced an interpolated function for local aspect ratio between thin and slim regimes, $h \equiv H/r = 0.03 + (h_{\rm slim}-0.03)\tanh(\mdot_{\rm loc})$, with $h_{\rm slim}=0.45$ as a fiducial value and $\mdot_{\rm loc}=\Mdot(r)/\Mdot_{\Edd}$ the local Eddington ratio. With radial viscous drift speed $v_r = \alpha h^{2} v_K$ (viscosity parameter $\alpha=0.1$ and $v_K=\sqrt{GM_{\rm BH}/r}$), the three governing timescales are given by:
\begin{equation}
t_{\rm adv} = \frac{r}{v_r} = \frac{t_{\rm dyn}}{\alpha h^{2}},\qquad
t_{\rm diff} = \frac{\tau_{\rm es} H}{c},\qquad
\tau_{\rm es} = \tfrac{1}{2}\kappa_{\rm es}\Sigma ,
\label{eq:timescales}
\end{equation}
with surface density $\Sigma = \Mdot(r)/2\pi r v_r$ from continuity. Additionally, $t_{\rm adv}$, $t_{\rm dyn}$, and $t_{\rm diff}$ correspond to the advection (inflow) time, local orbital (dynamical) time, and photon-diffusion time, respectively; $\tau_{\rm es}$ is electron-scattering optical depth with $\kappa_{\rm es}=0.34$ in CGS units. We assumed that the radiated flux is the viscous dissipation capped by the advective limit, $F_{\rm rad} = F_{\rm visc}\min(1, t_{\rm adv}/t_{\rm diff})$. This assumption swiftly reproduces the slim-disk temperature saturation ($T_{\rm eff}\propto r^{-1/2}$ inside the trapping radius).

Following above assumption, the bolometric luminosity, $L_{\rm bol}=\int 4\pi r F_{\rm rad}\,dr$, saturates as follows: $L_{\rm bol}/L\Edd\simeq2.4$, 5.4, 8.8, and 13 at $\mdot=3$, 10, 30, and 100, which is somewhat similar to the canonical $L\simeq L\Edd(1+\ln\mdot)$ \citep{ss73, poutanen07}, because the disk outside $r_{\rm sph}$ radiates its full dissipation. Throughout this work, $\mdot$ is the primary quantity to be constrained; we quote the Eddington Luminosity ratio $\lambda\Edd=L_{\rm bol}/L\Edd$ only through this explicit conversion, since $\lambda\Edd\neq\mdot$ once the luminosity saturates.

We define the trapping radius by the position where $t_{\rm diff}=t_{\rm adv}$. Substituting $\Sigma$ from continuity into Eq.~(\ref{eq:timescales}), we obtain:
\begin{equation}
\frac{t_{\rm diff}}{t_{\rm adv}} = \frac{\kappa_{\rm es}\,
\Mdot(r)\,h}{4\pi r c} \simeq 8.5\,\mdot_{\rm loc}\,h\,\frac{\rg}{r},
\label{eq:trapratio}
\end{equation}
where we uses $L\Edd=4\pi GMm_{\rm p}c/\sigma_{\rm T}$ with $\Mdot\Edd=L\Edd/0.1c^{2}$. Here, $\sigma_{\rm T}$ and $m_{\rm p}$ are the Thomson scattering cross-section for electron and mass of the proton. By setting this ratio to unity with $h\rightarrow h_{\rm slim}=0.45$ (valid for $\mdot_{\rm loc}\gg1$) gives:
\begin{equation}
r\trap \simeq 8.5\,h_{\rm slim}\,\mdot\,\rg \simeq 3.8\,\mdot\,\rg
\qquad (\mdot\gtrsim5),
\end{equation}
which is independent of $M_{\rm BH}$ and $\alpha$. Finally, the wind photosphere radius can be written as $r_{\rm ph} = \kappa_{\rm es}\Mdot_{\rm w}/4\pi v_{\rm w}$ with $v_{\rm w}\simeq v_{\rm esc}(r_{\rm sph})$, scales as $r_{\rm ph}\sim {\rm few}\times10^{2}\,\rg\,(\mdot/10)^{3/2}$, where $v_{\rm esc}$ is the local escape velocity.

A dissipation fluctuation born at radius $r$ is observed only if it survives damping mechanisms present. Here, we consider such three most appropriate mechanisms. (i) Advective attenuation: inside the trapping region, only a fraction $t_{\rm adv}/t_{\rm diff}$ of a local luminosity perturbation escapes before being advected onto the black hole,
\begin{equation}
A_{\rm adv}(r) = \min\!\left(1,\; t_{\rm adv}/t_{\rm diff}\right);
\end{equation}
(ii) Photospheric averaging: photons last scatter at $r_{\rm ph}$, whose surface contains $N \approx 2(r_{\rm ph}/H)^{2}$ independent coherence patches. The photospheric area $4\pi r_{\rm ph}^{2}$ divided into patches of area $\sim$$2\pi H^{2}$, since diffusion at the last-scattering surface spatially averages fluctuations over the local scale height, and due to that the emission from $r<r_{\rm ph}$ is correspondingly attenuated by a factor
\begin{equation}
A_{\rm ph} = N^{-k},\qquad k\in\{\tfrac12,\,1,\,\tfrac32\}.
\end{equation}
Since the averaging law is unknown, we bracket it with three exponents (central-limit, linear, and steep). (iii) Diffusion low-pass filter: photons random-walking out of radius $r$ over photon-diffusion time $t_{\rm diff}$ cannot transmit fluctuation power at Fourier frequencies $f$ above $(2\pi t_{\rm diff})^{-1}$ \cite[for discussion see][]{Kelly11}. Therefore, the transfer function for the photons can be written as
\begin{equation}
G(f,r) = \left[1 + (2\pi f\,t_{\rm diff})^{2}\right]^{-1/2}.
\end{equation}
Integrating $G^{2}$ against the Lorentzian power spectrum of damped random walk (DRW) of a zone, i.e., ${\rm PSD}(f)\propto[1+(2\pi f t_{\rm visc})^{2}]^{-1}$, gives the closed-form variance factor,
\begin{equation}
G_z^{2} = \frac{t_{{\rm visc},z}}{t_{{\rm visc},z} + t_{{\rm diff},z}}.
\label{eq:gzone}
\end{equation}
This expression correctly reduces to unity for slowly fluctuating zones ($t_{\rm visc}\gg t_{\rm diff}$) regardless of the observing baseline. The subscript 'z' stands for the value of the quantity at the radial extent of a zone.

The most important quantity for this work is to calculate a variability measure, which we estimate following the fluctuating-disk framework \citep{lyubarskii97, arevalo06}. To do that we divide the accretion disk into radial coherence zones of width $\Delta\ln r = 2h$.

Each zone undergoes a DRW with coherence time equal to its local inflow time, contributing to the band at wavelength $\lambda$ with weight $w_z$ set by the Planck function at $T_{\rm eff}(r)$. The observed fractional variance on a rest-frame baseline $\Delta t$ is
\begin{equation}
\begin{aligned}
& \sigma_m^{2}(\mdot, M_{\rm BH}, \lambda,\Delta t)  =  \\ & \sum_z
\left[w_z\,\sigma_0\sqrt{1-e^{-\Delta t/t_{{\rm visc},z}}}\,
\langle A_{\rm adv}A_{\rm ph}G\rangle_z\right]^{2},
\end{aligned}
\label{eq:master}
\end{equation}
i.e., DRW amplitude is suppressed by its flux-weighted damping factors (the full derivation is given in Appendix~\ref{app:master}). We calibrate the single free amplitude $(\sigma_0)$ to reproduce the observed fractional rms of a fiducial sub-Eddington quasar, providing an empirical normalization for the variability model. The value of fractional rms $\sigma_m = 0.15$ for an ordinary quasar ($M=10^{8}\,M_\odot$, $\mdot=0.3$, $\lambda=2500$\,\AA, $\Delta t=1$\,yr; \citealt{macleod10, kelly09, wilhite08, burke21}), giving us the value of $\sigma_0 \simeq 2.1$ per zone, where $M_\odot$ is the solar mass.

\section{Results} \label{sec:results}
\begin{figure}[t]
\centering
\includegraphics[width=\columnwidth]{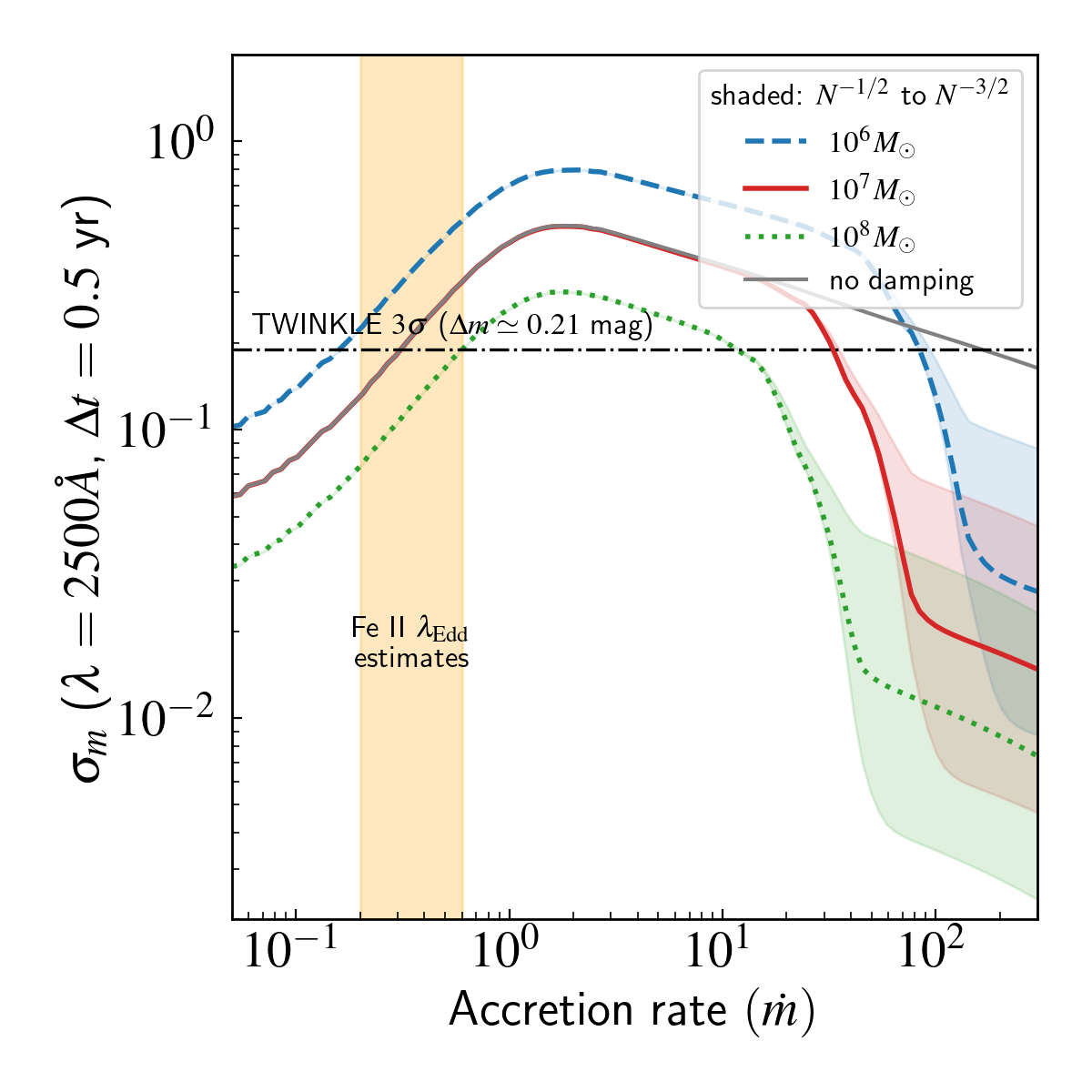}
\caption{The maximum observable rms fractional variability ($\sigma_m$) at rest with $\lambda=2500\,\AA$, $\Delta t=0.5$\,yr as a function of accretion rate ($\mdot$) for three black hole masses. Shaded bands span photospheric-averaging laws $N^{-1/2}$ to $N^{-3/2}$ (lines: $N^{-1}$). The horizontal line marks the TWINKLE continuum (dot-dashed) detection threshold; the gray curve shows the undamped thin-disk reference; the orange band marks Fe\,{\sc ii}-based sub-Eddington estimates ($\mdot\simeq\lambda\Edd\simeq0.2$--$0.6$ for $\eta=0.1$), which fall in the detectable region.}
\label{fig:ceiling}
\end{figure}

Figure~\ref{fig:ceiling} shows $\sigma_m(\mdot)$ at rest with $\lambda=2500\,\AA\ $ and $\Delta t=0.5$\,yr (matching the TWINKLE baseline). Below $\mdot\sim1$ the curves track the undamped thin-disk behavior. Above $\mdot\sim{\rm a~few}$ the three mechanisms compound, and the rms fractional variability deviates and collapses by orders of magnitude. For example, $M_{\rm BH}=10^{7}\,M_\odot$ curve is crossing the TWINKLE continuum $3\sigma$ detection threshold ($\Delta m\simeq0.21$\,mag, $\sigma_m\simeq0.19$ adopting the JWST $3\sigma$ convention of \cite{Zhou-etal2025,secunda26}) at $\mdot \simeq 30$. Note that $\Delta m$ is the magnitude rms corresponding to the fractional rms $\sigma_m$, they are related by $\Delta m= (2.5/\ln{10})\sigma_m$. Thus the TWINKLE non-detections therefore imply
\begin{equation}
\mdot \gtrsim 30 \qquad (M_{\rm BH}\sim10^{7}\,M_\odot,\; \Delta t = 0.5\,{\rm yr}),
\end{equation}
or $\lambda\Edd\gtrsim9$ with the model luminosity ($\gtrsim4$ for $L=L\Edd(1+\ln\mdot)$). Considering the accreting black hole interpretation adopted in this work, the sub-Eddington values $\lambda_{\rm Edd}\simeq0.2–0.6$ inferred from Fe II-based calibrations imply $\dot m\simeq0.2–0.6$ for standard efficiency. These values lie deep in the detectable region, making them incompatible with the observed lack of continuum variability. Additionally, we find that the limit tightens with monitoring baseline by a factor of $\approx3$ per decade in $\Delta t$ (crossing at $\mdot\simeq15$, 30, 50 for $\Delta t=0.15$, 0.5, 2\,yr). This suggests that longer campaigns probe deeper into the ceiling. However, the value saturates with further increasing of $\Delta t$ to $\mdot_{\rm min}\simeq85$.

\begin{figure*}[t]
\centering
\includegraphics[width=0.95\textwidth]{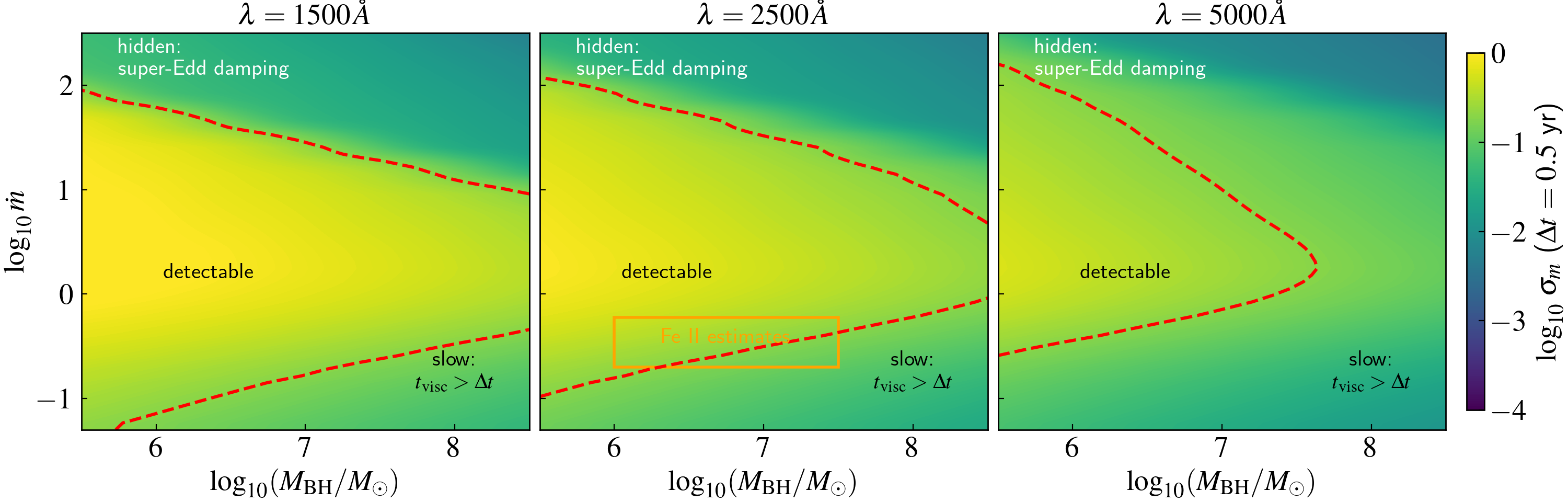}
\caption{Detectability maps in the $(M_{\rm BH},\mdot)$ plane at three wavelengths (1500, 2500, 5000\,\AA; $\Delta t=0.5$\,yr). The red contour is the TWINKLE $3\sigma$ threshold. The orange box marks Fe\,{\sc ii}-based sub-Eddington estimates ($\mdot\simeq\lambda\Edd$ for $\eta=0.1$).}
\label{fig:map}
\end{figure*}

The characteristic timescales for lower-mass black holes are shorter; therefore, they vary more and are harder to hide. For example, at $10^{6}\,M_\odot$ the upper limit requires $\mdot\gtrsim100$, while $10^{8}\,M_\odot$ reaches below the limits at $\mdot\sim10$ (Figures~\ref{fig:ceiling} and \ref{fig:map}). The detectability maps of Figure~\ref{fig:map} divide the $(M_{\rm BH},\mdot)$ plane by the TWINKLE threshold, which is shown at three rest-frame wavelengths (1500, 2500, and 5000\,\AA). The boundary shifts to lower $\mdot$ at longer wavelengths, so longer wavelength bands hide high-$\mdot$ sources more readily. Note that high-mass but low-$\mdot$ (ordinary AGNs) are hidden on the map not because of super-Eddington damping but due to longer viscous timescales than those of the baseline $(t_{\rm visc}>\Delta t)$. Thus, even an ordinary AGN may appear constant on the TWINKLE baseline. This suggests that the  non-detections of $M_{\rm BH}\gtrsim10^{8}\,M_\odot$ LRDs merely cannot constrain $\mdot$ unless the baseline is long enough that $\mdot_{\rm min}$ falls within reach (as discussed in the previous paragraph). Therefore the observed variable minority of LRDs ($\sim$3\%; \citealt{zhang25}, and possibly lensed systems monitored over longer effective baselines) should then be systematically lower-$\mdot$ and/or more massive than the non-variable population. In principle, this argument is testable with existing samples.

Next, we forward model the TWINKLE campaign for 18 LRDs with baselines of $\Delta t=$140--220 days sampled by 12 epochs \citep{liu26}. We draw masses log-uniformly from $10^{6.5}$--$10^{8}\,M_\odot$, approximating the commonly assumed mass range \citep{liu26}, and adopt a per-epoch photometric precision of $0.07$\,mag, following the $3\sigma$ threshold convention of \citet{secunda26}; both the mass distribution and the photometric precision are varied in Table~\ref{tab:mcsens}. For each source the model supplies not merely an rms value but the full second-order statistics of its light curve. The zone decomposition of Section~\ref{sec:math} (derived in Appendix~\ref{app:master}) represents the band flux as a sum of independent DRWs, one per coherence zone, with per-zone variances $v_z$ and coherence times $\tau_z$. Equivalently, the light curve is a Gaussian process with covariance $C(\Delta t)=\sum_z v_z\,e^{-\Delta t/\tau_z}$. We draw realizations of this process at the source's actual twelve epochs by Cholesky factorization of the epoch-to-epoch covariance matrix, where we add Gaussian photometric noise of $0.07$\,mag per epoch. We count the source as detected when the $\chi^2$ of the sampled light curve about its own mean exceeds the $3\sigma$ threshold for $n_{\rm ep}-1$ degrees of freedom. Repeating this process over $2\times10^{4}$ realizations yields the per-source detection probability $p_{\rm det}(M_{\rm BH},\mdot,\Delta t)$. Finally, averaging $\prod_i(1-p_{{\rm det},i})$ over the mass and baseline distributions of the 18 sources gives the probability of zero detections, $P(N_{\rm det}=0\mid\mdot)$, which is shown in Figure~\ref{fig:mc} ({\it Left}) (full details in Appendix~\ref{app:mc}) as a function of accretion rate for two wavelengths $\lambda=2500, 5000\,\AA$. The figure suggests that the $95\%$ confidence lower limit is $\dot m \gtrsim 60$ at $2500 \AA$ and $\dot m \gtrsim 70$ at $5000 \AA$, a factor of two stronger than the simple threshold comparison of Figure~\ref{fig:ceiling}. This is because with twelve epochs, even an rms at the single-epoch threshold is almost certainly detected somewhere in the light curve. For the fiducial mass distribution, the limit is insensitive to cadence (6–20 epochs) and only weakly dependent on the photometric precision ($\mdot\gtrsim50$--70 for $0.1-0.05$\,mag). Similarly, for different plausible mass distributions, it ranges from $\mdot\gtrsim40$ ($10^{7}$--$10^{8}\,M_\odot$) to $\gtrsim100$ ($10^{6}$--$3\times10^{7}\,M_\odot$); thus, the dominant uncertainty is the LRD mass function, not the statistics (see Table~\ref{tab:mcsens} for details). Additionally, it is worth mentioning that the high-$\mdot$ plateau at $P\simeq0.95$ reflects a false-alarm rate of 0.27\% per source.

\begin{figure*}[t]
\centering
\includegraphics[width=0.8\columnwidth]{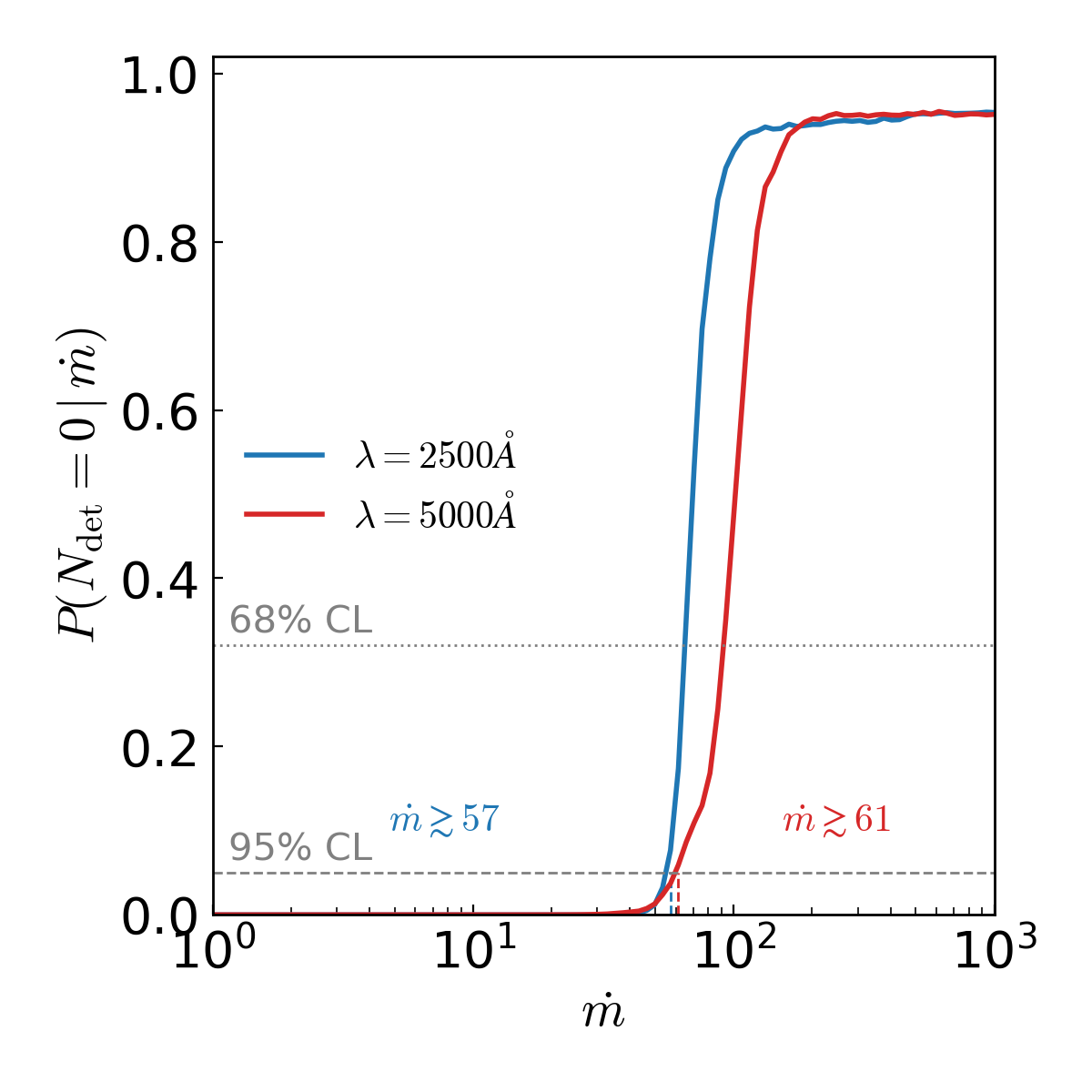}
\includegraphics[width=0.8\columnwidth]{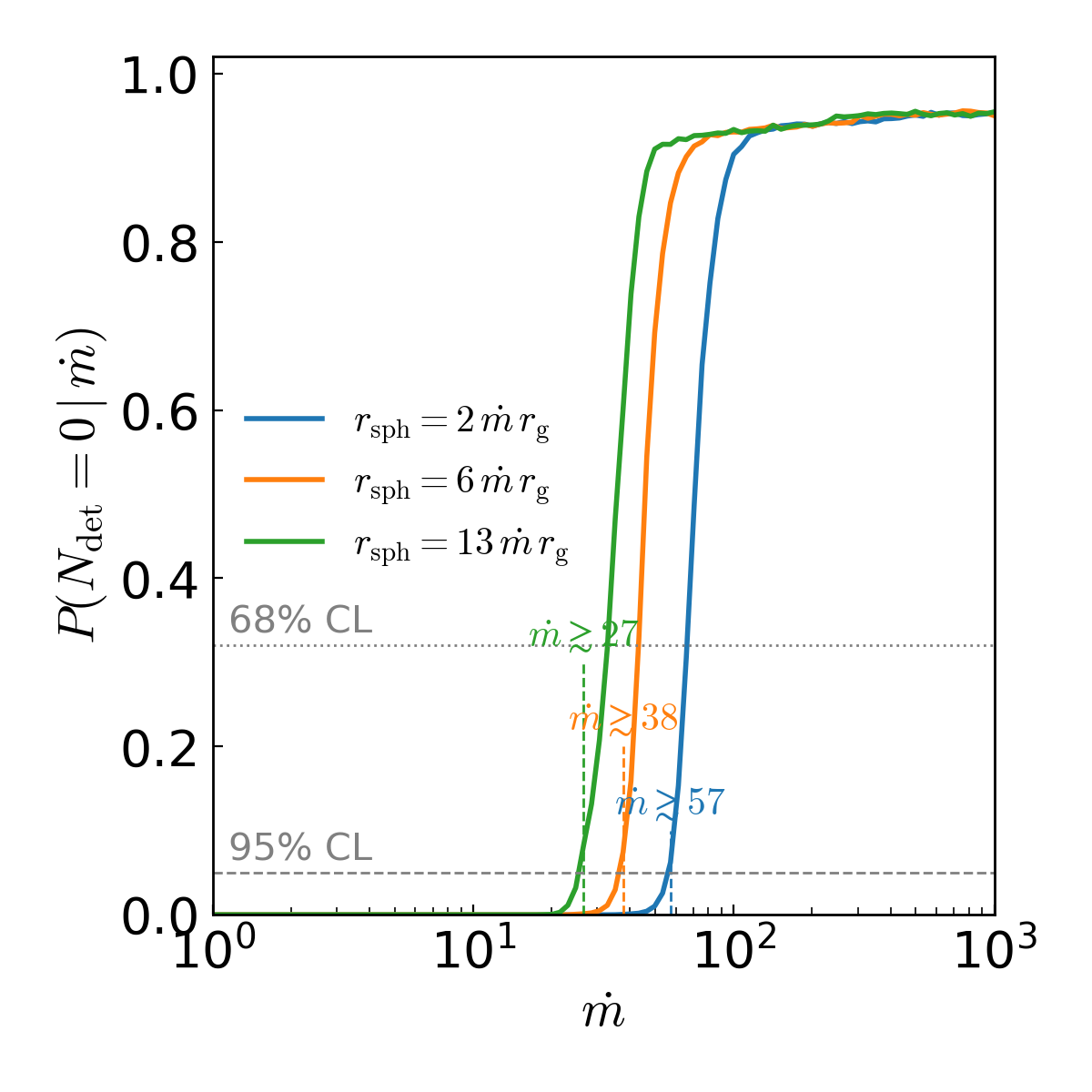}
\caption{{\it Left:} Probability of detecting zero variable sources among 18 TWINKLE LRDs as a function of the accretion rate for $\lambda=2500\AA$ and $5000\AA$, from the Monte Carlo forward model (Appendix~\ref{app:mc}). {\it Right:} Same as the {\it Left} but for different $r_{\rm sph}$.}
\label{fig:mc}
\end{figure*}

Since the governing microphysics are the same for both ordinary quasars and LRDs. Therefore, the model must be able to be recalibrated for each parameter choice and, accordingly $\sigma_0$ is refitted to the ordinary quasars for every $(\alpha, h_{\rm slim})$ combination. These studies reveal that the accretion rate threshold is invariant within $\mdot\simeq29$--$35$ for $h_{\rm slim}=0.35$--$0.8$ and $\alpha=0.03$--$0.3$ (See Appendix~\ref{app:calib} for more details). On the contrary, the dominant model systematic is the wind mass-loss law; with the removal of wind completely ($s=0$), the threshold on $\mdot$ rises to $\mdot\simeq120$.

Likewise, by adopting $r_{\rm sph}=6$--$13\,\mdot\,\rg$ instead of $2\,\mdot\,\rg$, the threshold moves to $\mdot_{\min}\simeq18$--23 (more wind mass, larger $r_{\rm ph}$; See Appendix~\ref{app:calib}). Accordingly, the 95\% confidence limit relaxes correspondingly from $\mdot\gtrsim60$ to $\mdot\gtrsim26$--38 Figure~\ref{fig:mc} ({\it Right}). We therefore emphasize that $\mdot_{\min}\simeq30$ is the fiducial prediction of the adopted SS73 wind model, which is not a universal critical value for all super-Eddington disks. What is universal in our results is the limit itself, the steep collapse of $\sigma_m$ above $\mdot\sim$ a~few to $10\,\mdot$, while its exact numerical threshold may differ slightly with the wind prescription.

\section{Discussion} \label{sec:disc}

Qualitatively, our conclusions are expected from previous studies, e.g., \citet{zhang25} invoked super-Eddington accretion for their non-variable majority of LRDs. Additionally, due to the photon trapping in the inner flow, geometrically thick accretion flow is expected to have weak UV/optical variability \citep{inayoshi25c,madau26} . The best quantitative comparison of our work can be done with \citet{secunda26}; they concluded with radiation-MHD simulations spanning $\Mdot=0.03$--$3\,\Mdot\Edd$ that $\mdot\sim{\rm a~few}$ could explain the non-detections. However, our estimations push the limits deeper. The discrepancy is understandable because their light curves retain only locally generated turbulence, which is extrapolated from a single black hole mass simulation, whereas our normalization inherits the observed rms of sub-Eddington quasars, including fluctuations propagating inward from large radii, and finally, applies damping on top.

The model predicts the variability magnitude in the rest frame by construction. The redshift enters only through the band mapping $\lambda_{\rm rest}=\lambda_{\rm obs}/(1+\Tilde{z})$ and the time-dilated baseline, so it can be tested directly against local accretion sources by evaluating at the observed bands and baselines. The details can be found in Appendix~\ref{app:local}. The model can roughly reproduce the observed trend of decreasing variability with increasing $\mdot$ across the major accretor classes, and it is quantitatively consistent with the super-Eddington population, the same regime relevant to the LRDs. In this regime the predicted rms falls to the few-percent level measured for the highest-$\dot m$ sources of \citet{lu19}. We deliberately do not claim agreement at sub-Eddington rates, where radiative-inefficient accretion flow (RIAF) models are more relevant than that of our model \citep{yuan14,Narayan1994Advection-dominatedSolution}.

Figure~\ref{fig:ceiling} shows that the LRD non-detections are formally degenerate with the standard sub-Eddington accretion regime, $\mdot\lesssim0.1$. Whether this branch represents a physically distinct alternative for LRDs or merely a formal degeneracy cannot be decided with the present observed baseline. Only a longer monitoring baseline will be able to discriminate between these two regimes. The apparent lack of variability in this regime has a different physical origin from the super-Eddington collapse, namely freezing rather than damping (see the gray solid curve of Fig.~\ref{fig:ceiling}). The frozen thin disk is quiet only on short baselines, as the inflow times far exceed the observing window, so the predicted rms grows with baseline (e.g., by a factor of $\sim3$ between 1 and 10\,yr at $\mdot\sim10^{-2}$; Appendix~\ref{app:local}). On the other hand, the super-Eddington collapse is almost baseline-independent (see Section~\ref{sec:results}). We therefore regard $\mdot\lesssim0.1$ as a formal degeneracy, which can be directly falsified with longer monitoring. Note that this discussion rests on the variability timescales alone. This neither requires the absolute bolometric luminosity scale of LRDs, which carries orderc of magnitude systematic uncertainty  \citep[e.g.,][]{Sacchi25,Greene2026}, nor any particular interpretation of the red continuum \citep[e.g.,][]{Delvecchio2025}.

One of the major questions is what will be the consequences for our framework for X-ray variability. The averaging exponent $k$ is nearly degenerate in the UV/optical (Figure~\ref{fig:ceiling}), where $r_\lambda \gtrsim r_{\rm ph}$ except at extreme $\mdot$ (where $r_\lambda$ is the flux-weighted emission radius of the band), but coronal X-rays which could originate at $r\sim{\rm few}\,\rg \ll r_{\rm ph}$ at all super-Eddington rates. So, we can expect that X-ray variability should be suppressed by all three mechanisms at once, with a strength that depends on $k$. Thus, the observed X-ray weakness of LRDs \citep{ananna24, kokubo25} is qualitatively consistent with this picture \citep[see also][]{inayoshi25c}. Additionally, the $\sim$3\% of LRDs with detected variability \citep{zhang25} can be naturally explained with low-$\mdot$ accretion flow or with the high-mass boundary of Figure~\ref{fig:map}. Their amplitudes ($\Delta m\sim0.3$--$0.8$\,mag) and short (observer-frame 10--100\,day) timescales, however, these values lie at the upper edge of what any disk model can produce. If they are further confirmed, they may indicate a different origin, e.g., transients, lensing, or quasi-star pulsations \citep{cantiello25} rather than the low-$\mdot$ tail of the disk population.

Within the super-Eddington picture, LRDs may be embedded in dense gas cocoons, which would further dilute the disk variability \citep{naidu25, degraaf25}. Note that these cocoon models for LRDs do not themselves require super-Eddington accretion scenario. Regarding the cocoons, two features of our model are relevant here. (i) Photospheric averaging in our model already acts as the self-consistent inner cocoon developed by the wind. This photosphere ($r_{\rm ph}$) acts as many fluctuating patches, which provides precisely the geometric dilution an extended envelope would offer. (ii) Any additional external envelope with a photosphere is nearly gray. The electron scattering conserves photon number, so the envelope cannot destroy variability power; however, it could redistribute it in arrival time. It suppresses variability solely through temporal smearing on $t_{\rm env}\sim(r_{\rm env}/c)\,\tau_{\rm es}\sim0.01$--0.2\,yr (for $r_{\rm env}\sim 10^3-10^4\,r_g (\sim10^{16}$\,cm for $10^7M_\odot$) and $\tau_{\rm es}\sim1$--10). Fluctuations faster than $t_{\rm env}$ are attenuated, while slower variations pass through with their amplitude. Taken together with the photospheric averaging already present in our model, an additional envelope may therefore marginally reduce the rms on the shortest baselines ($\propto \sqrt{1-t_{\rm env}/\Delta t}\,\Rightarrow$ at most a factor of $\sim2$), but its effect is expected to vanish on longer baselines. Note that more extended or optically thicker envelopes, or envelopes involving substantial absorption and reprocessing, could produce stronger suppression and are beyond the scope of the present work.

In this work, the model is designed to be minimal. We neglect (i) re-emergence of advected heat, which would slightly reduce the boundary of $\mdot_{\rm min}$; (ii) general-relativistic corrections and (iii) spin, which shifts inner radii; (iv) H$\alpha$ variability, which involves the BLR response and lies beyond the continuum framework; and (v) the per-zone amplitude $\sigma_0$, which is not predicted or derived but is calibrated to the observed rms of ordinary quasars. We also flag (vi) our assumption that each zone fluctuates on its local inflow time: for the calibration quasar ($10^{8}\,M_\odot$, $\mdot=0.3$) this predicts a variability timescale of $\approx$20--30\,yr (scaling as $M^{1/2}$) at 2500\,\AA\, an order of magnitude longer than observed quasar timescales ($\sim$1--2\,yr), which is a well-known, generic shortcoming of viscous-time fluctuation models. In our work, the $\sigma_0$ is recalibrated separately for all the parameter sets. Thus, the threshold depends on timescale ratios rather than absolute values, which is why it is invariant under $h_{\rm slim}$ and $\alpha$ variations (Figure~\ref{fig:hslim}). Additionally, in our model, the outer flow remains thick ($h \simeq h_{\rm slim}$) at all radii, supported by a super-Eddington wind-fed rate \citep{poutanen07}. Radiation-hydrodynamic and radiation-MHD simulations of super-critical accretion consistently find geometrically thick inflows well beyond $r_{\rm sph}$ \citep[e.g.,][]{ohsuga05,jiang19,sadowski16}. Moreover, much of the observed UV continuum emerges inside the wind photosphere $r_{\rm ph}$ (see Section~\ref{sec:results}), where the flow is thick by construction. Note that, where the predicted rms exceeds unity (low-mass corner of Figure~\ref{fig:map}) the model should be read as saturated at the coherent-fluctuation limit, and our conclusions do not depend on that regime. Finally, none of these caveats affects the central result: the limits on the $\mdot$ are set by $r\trap\propto\mdot$, common to all supercritical disk models.

\section{Conclusions} \label{sec:conc}

We have shown that super-Eddington accretion disks possess a maximum observable variability due to photon trapping, photospheric averaging, and diffusion-time filtering, which suppress continuum variability by orders of magnitude above $\mdot\sim{\rm a~few}$. Applying our models to JWST's LRDs, the TWINKLE non-detections require $\mdot\gtrsim30$ at $10^{7}\,M_\odot$ ($\mdot\gtrsim60$ at 95\% confidence in our forward model) and $\mdot\gtrsim100$ at $10^{6}\,M_\odot$. However, the precise threshold is tied to the wind prescription ($\mdot_{\min}\simeq30$ fiducial, $\simeq120$ wind-free); the suppression of variability itself is not, even with no wind at all. The photon trapping alone forces $\sigma_m$ to decline with $\mdot$, since $r\trap\propto\mdot$ is common to all supercritical disk models. On the other hand, a frozen sub-Eddington thin disk at $\mdot\lesssim0.1$ remains formally degenerate with the non-detections on current baselines. The predicted variability in this regime grows with the baseline, unlike the super-Eddington collapse, which is nearly baseline-independent. So longer monitoring will directly discriminate between these two regimes. The variable minority is probably mass-segregated or driven by different mechanisms. If future deep monitoring instead reveals month-scale variability across the LRD population, super-Eddington disk models can be falsified. Either way, these results are useful. If LRDs are variable, the variability measures the accretion rate; if they do not, the non-detections favor the super-Eddington branch within this framework, while the frozen sub-Eddington alternative remains directly falsifiable with longer baselines.

\begin{acknowledgments}
The author thanks the anonymous referee for a careful and constructive report, which substantially improved the manuscript. I am deeply grateful to my wife, Jharna, and our son, little Ivaant, for their love, patience, and quiet sacrifices while we remain apart. The author used an AI assistant in developing the analysis Python scripts and editing the manuscript and takes full responsibility for its content. This work is supported by the National Natural Science Foundation of China grant no. W2633017 and the Ningbo Science \& Technology Bureau program grant No.2024DJRC010009.
\end{acknowledgments}

\section*{Data Availability}
The scripts and model are publicly available online at DOI:10.5281/zenodo.23968137.


\appendix
\twocolumngrid
\section{Variability estimator} \label{app:master}

The results of this work are based on the variability estimator Eq.~(\ref{eq:master}). So, here, we collect the steps to getting this equation. For a face-on disk, the observed specific flux in a band centered on $\lambda$ is $F_\lambda \propto \int 2\pi r\, B_\lambda[T_{\rm eff}(r)]\,dr$. The advantage of fractional rms is that all constants cancel out in the final expression. So we use normalized radial weights $w(r) = 2\pi r\,B_\lambda[T_{\rm eff}(r)]/\int 2\pi r'\, B_\lambda[T_{\rm eff}(r')]\,dr'$ with $\int w\,dr = 1$. These weights tell us which radii the band actually sees. Next we calculate how each contributing region fluctuates.

For that we adopt the fluctuating-disk picture \citep{lyubarskii97, arevalo06}. In the fluctuating-disk framework, a surface-density perturbation is advected inward with the local inflow time and remains coherent over a radial range set by the disk scale height ($H$). We therefore divide the disk into connected logarithmic zones $[r_z,\, r_z e^{2h(r_z)}]$. Each zone $z$ carries a fraction $f_z = \int_z w(r)\,dr$ of the band flux ($\sum_z f_z = 1$) and fluctuates on the geometric-mean inflow time $\tau_z = \exp\langle\ln t_{\rm visc}\rangle_z$, where $\langle ~\,\rangle$ denotes the zonal average of a quantity. Having defined the zones and their timescale, we now specify the stochastic process for each zone.

The fractional luminosity fluctuation of each zone $\delta_z$ obeys a Langevin equation with stationary variance $\sigma_0^{2}$ and covariance $\langle\delta_z(t)\,\delta_z(t+\Delta t)\rangle = \sigma_0^{2}\, e^{-\Delta t/\tau_z}$. Thus, a campaign of duration $\Delta t$ measures fluctuations about the campaign mean, with variance:
\begin{equation}
\sigma_z^{2}(\Delta t) = \tfrac{1}{2}\left\langle
\left[\delta_z(t+\Delta t) - \delta_z(t)\right]^{2}\right\rangle
= \sigma_0^{2}\left(1 - e^{-\Delta t/\tau_z}\right),
\label{eq:drw}
\end{equation}
i.e., half the two-epoch structure functions ($SF$). It is easy to realize that the variance reduces to $\sigma_0^{2}$ for $\Delta t\gg\tau_z$ and to $\sigma_0^{2}\,\Delta t/\tau_z$ for $\Delta t\ll\tau_z$. This is the variance one would measure if photons escaped the disk freely. But in a super-Eddington disk they cannot escape freely. The first obstacle to free escape is photon diffusion. The intrinsic fluctuation is convolved with the photon-diffusion propagator, multiplying its power spectrum by $G^{2}(f) = [1+(2\pi f\,t_{\rm diff})^{2}]^{-1}$. With the two-sided DRW spectrum $P(f) = 2\sigma_z^{2}\tau_z/[1+(2\pi f\tau_z)^{2}]$ and $\int_{-\infty}^{\infty}df\,[(1+(2\pi f\tau)^{2}) (1+(2\pi f\,t_{\rm diff})^{2})]^{-1} = [2(\tau+t_{\rm diff})]^{-1}$, the observed variance is $\sigma_{z,\rm obs}^{2} = \sigma_z^{2}\,\tau_z/(\tau_z + t_{{\rm diff},z})$, which is Eq.~(\ref{eq:gzone}). The factor $\tau_z/(\tau_z + t_{{\rm diff},z})$ approaches unity for $t_{\rm diff}\ll\tau_z$ and $\tau_z/t_{\rm diff}$ in the opposite limit. Diffusion is not the only attenuation channel: advection into the black hole and averaging over photospheric patches act alongside it. Therefore, all three of these must be applied together.

We therefore combine the three damping factors into the zone fluctuations before summing. The damping factors vary slowly within a zone, so each zone carries its flux-weighted mean $A_z = f_z^{-1}\int_z w(r)\,A_{\rm adv}A_{\rm ph}G\,dr$. The observed fractional fluctuation is $\delta F/F = \sum_z f_z A_z\,\delta_z$; because the zones are statistically independent by construction; therefore, the cross terms vanish. This finally yields the variability estimator, i.e., Eq.~(\ref{eq:master}).

\section{Recalibration of estimator} \label{app:calib}

\begin{figure*}[t]
\centering
\includegraphics[width=0.8\textwidth]{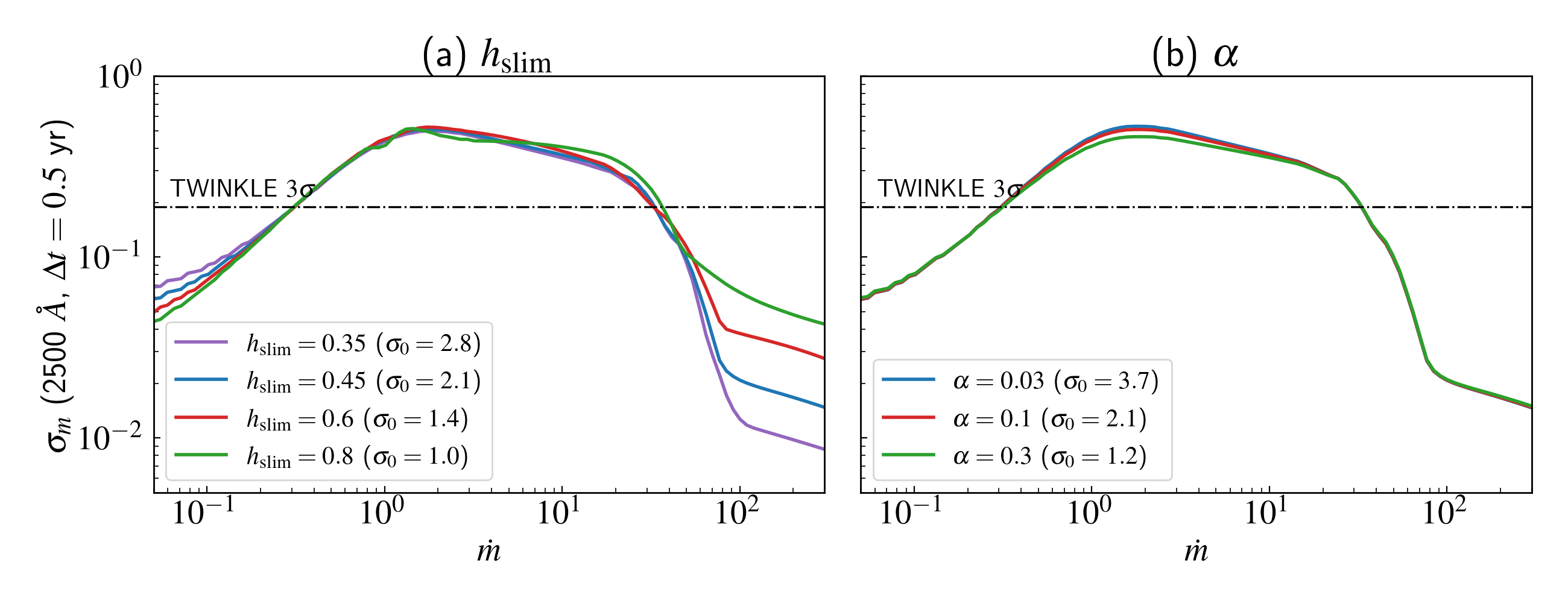}
\caption{ Same as Figure~\ref{fig:ceiling}, but shown for different (a) disk-height ($h_{\rm slim}$) and (b) viscosity parameter $(\alpha)$.}
\label{fig:hslim}
\end{figure*}

\begin{figure*}[t]
\centering
\includegraphics[width=0.80\textwidth]{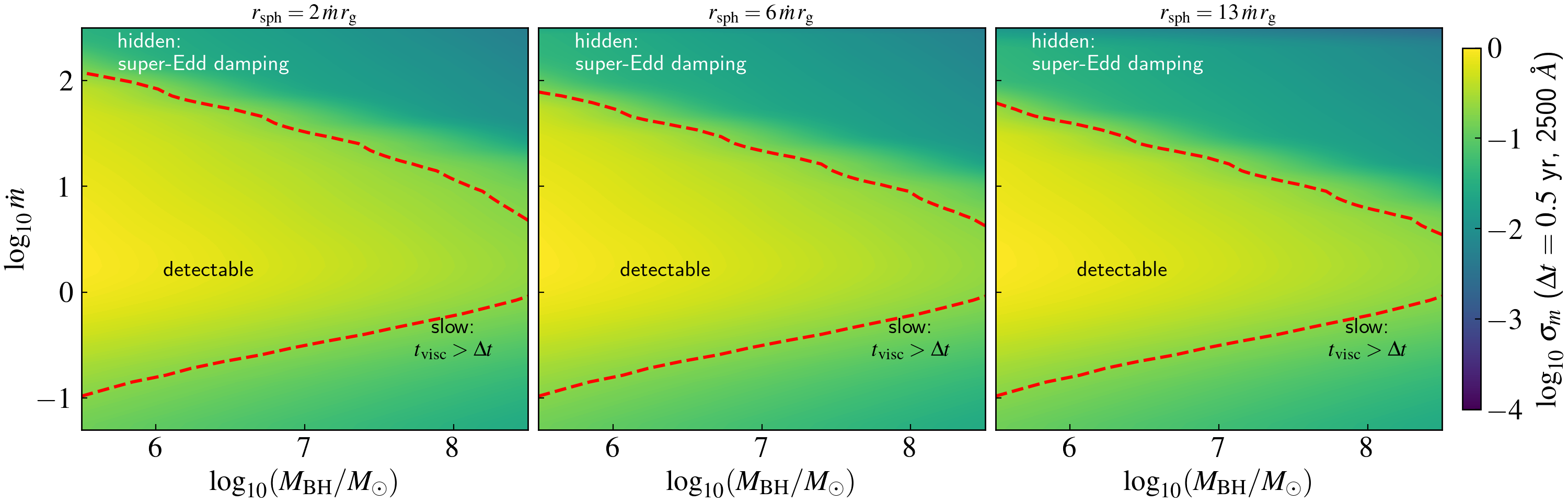}
\caption{Same as Figure~\ref{fig:map}, but for different spherization
radius, $r_{\rm sph}=2$, 6, and $13\,\mdot\,\rg$ (left to right).}
\label{fig:rsph}
\end{figure*}

In the panels of Figure~\ref{fig:hslim}, we show the robustness of the modeling by varying (a) disk height ($h_{\rm slim}$) and (b) viscosity parameter ($\alpha$). Each curve is recalibrated to the ordinary quasar anchor for its own parameter set and corresponding $\sigma_0$ given on the legends. The parameter dependence cancels between calibration and limits of accretion rate ($\mdot$). Thus, the threshold at $\mdot\simeq30$ is invariant across different parameters. We observe that the curves separate only deep in the hidden regime. Similarly, in Figure \ref{fig:rsph}, we have shown the detectability maps at $\lambda=$2500\,\AA\ (as in Figure~\ref{fig:map}) for three normalizations of the spherization radius, $r_{\rm sph}=2$, 6, and $13\,\mdot\,\rg$ (left to right). These span the range used in the earlier supercritical-disk literature \citep{ss73, abramowicz88, poutanen07}. A larger $r_{\rm sph}$ drives more mass through the wind and enlarges the wind photosphere. This strengthens both the trapping and photospheric-averaging damping. Thus the hidden/detectable boundary at $M=10^{7}\,M_\odot$ slides from $\mdot_{\min}\simeq34$ to 23 to 18, which is not drastic. Accordingly, in all cases the preference for the super-Eddington branch is unchanged, and the formal degeneracy with the frozen sub-Eddington branch (Section~\ref{sec:disc}) is also independent of these parameters.

\section{Monte Carlo forward model}
\label{app:mc}
\setcounter{equation}{0}

We model the TWINKLE LRD survey considering a sample consisting of $N_{\rm src}=18$ sources with black hole masses drawn log-uniformly from $10^{6.5}$--$10^{8}\,M_\odot$. Similarly, the rest-frame baselines are drawn uniformly from 140--220\,d, $n_{\rm ep}=12$ epochs spaced evenly over the baseline, and a per-epoch photometric precision $\sigma_{\rm phot}=0.07$\,mag (fractional $0.064$), corresponding to the quoted $3\sigma$ threshold $\Delta m\simeq0.21$\,mag. With these survey parameters, each source's light curve is generated as a Gaussian process. For a given $(M_{\rm BH},\mdot,\lambda)$, the band light curve is zero-mean with covariance:
\begin{equation}
C_{ij} = \sum_z v_z\,\exp\!\left(-|t_i-t_j|/\tau_z\right)
+ \sigma_{\rm phot}^{2}\,\delta_{ij},
\label{eq:cov}
\end{equation}
where $v_z$ and $\tau_z$ are the per-zone variance coefficients and coherence times of Appendix~\ref{app:master} (including all damping factors and the calibration $\sigma_0$), evaluated at the mass and accretion rate of the source. Finally, the synthetic light curves are drawn by Cholesky decomposition of $C_{ij}$.

These synthetic light curves are then passed through the same detection criterion applied to the real data. A source is counted as variable if $\chi^2 = \sum_i (x_i-\bar{x})^{2}/\sigma_{\rm phot}^{2}$ exceeds the $3\sigma$-equivalent threshold $\chi^2_{n_{\rm ep}-1}(1-0.0027)$ ($=27.2$ for $n_{\rm ep}=12$). The per-source detection probability $p_{\rm det}(M_{\rm BH},\mdot,\Delta t)$ is measured from $2\times10^{4}$ simulated light curves on a grid of 6 masses $\times$ 3 baselines. The per-source probabilities are then combined into a survey-level statistic: for each of $4\times10^{3}$ the population realizations we draw 18 (mass, baseline) pairs and compute $P_0 = \prod_i [1-p_{{\rm det},i}]$; the plotted $P(N_{\rm det}=0\mid\mdot)$ is the population average of $P_0$. The 95\% confidence lower limit is the smallest $\mdot$ with $P(N_{\rm det}=0)\geq0.05$. The high-$\mdot$ plateau, $P_0^{\rm max}=(1-0.0027)^{18}=0.952$, is set by the false-alarm rate of the $\chi^2$ selection. Finally, we test how sensitive the resulting limit is to the assumptions adopted above. Table~\ref{tab:mcsens} lists the 95\% limit under variations of the survey assumptions, where we observe that the dominant uncertainty arises due to the adopted mass distribution.

\begin{deluxetable}{lc}
\tablecaption{Sensitivity of the 95\% CL lower limit on $\mdot$
(at $\lambda=$2500\,\AA).\label{tab:mcsens}}
\tablehead{\colhead{Assumption varied} & \colhead{$\mdot_{\min}$ (95\% CL)}}
\startdata
Fiducial & 60 \\
$n_{\rm ep}=6$ & 60 \\
$n_{\rm ep}=20$ & 60 \\
$\sigma_{\rm phot}=0.05$\,mag & 70 \\
$\sigma_{\rm phot}=0.10$\,mag & 50 \\
$M_{\rm BH}\in10^{6}$--$3\times10^{7}\,M_\odot$ & 100 \\
$M_{\rm BH}\in10^{7}$--$10^{8}\,M_\odot$ & 40 \\
\enddata
\end{deluxetable}
\begin{figure*}[t]
\centering
\includegraphics[width=0.8\textwidth]{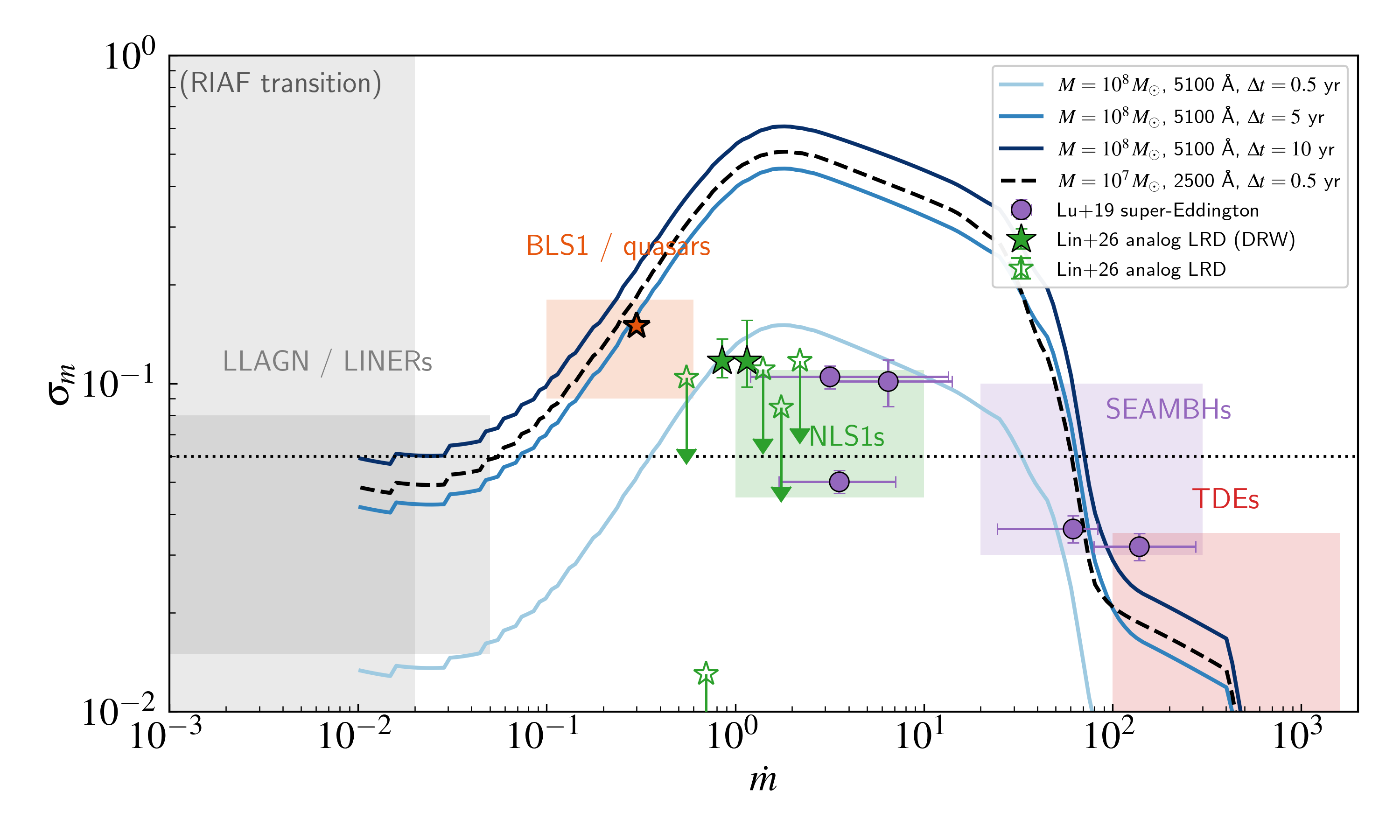}
\caption{Model $\sigma_m$ versus $\mdot$ for the local-accretor comparison. The horizontal bands mark the characteristic variability of each class; the star corresponds to the quasar calibration point. The gray region is the transition to RIAF, and the purple points are the reliable super-Eddington subsample of \citet{lu19}. Green stars correspond to data of analog LRDs from \cite{Lin26}: filled symbols are the two DRW detections and and open symbols with downward arrows are 90\% upper limits on $SF_\infty$(see Appendix~\ref{app:local} for more details).}
\label{fig:local}
\end{figure*}
\section{Comparison with local accretors} \label{app:local}

The framework developed here is naturally in rest-frame, so it can be used directly on nearby accretors. Figure~\ref{fig:local} shows a comparison of the predicted fractional rms at rest-frame 5100\,\AA\ ($M_{\rm BH}=10^{8}\,M_\odot$; baselines 0.5--10\,yr) and at rest-frame 2500\,\AA\ ($M_{\rm BH}=10^{7}\,M_\odot$; baseline 0.5\,yr) with the main classes of local accretor. The near-Eddington branch ($\mdot\simeq0.1$--$0.6$) matches broad-line Seyfert 1 galaxies (BLS1s) and quasars by construction, since the estimator is calibrated in this range of accretion rate \citep{macleod10}; all other regions are genuine predictions of our model.

The class-by-class behavior follows: (i) Narrow-line Seyfert 1 galaxies (NLS1s, $\mdot\sim1$--10): Within the NLS1 population itself,  the variability amplitude anti-correlates with the Eddington ratio \citep{ai10, rakshit17, wang22}, the trend our model reproduces. A least-squares fit from our model gives $\sigma_m\propto\mdot^{-0.12}$ over $\mdot\simeq1$--$10$ ($M=10^8\,M_\odot$, rest-frame 5100\,\AA, $\Delta t=5$\,yr), in close agreement with the ensemble slope $\sigma_d\equiv\sigma_m(\Delta t\gg\tau_d)\propto\dot M^{-0.13\pm0.03}$ measured by \citet{lu19} and with the independent DRW amplitude scaling $\sigma_{2500}\propto\mdot^{-0.13\pm0.02}$ measured from AGNs with baselines of up to 20\,yr \citep{ning26}. Note that here we only compared the trend, not the absolute values. (ii) Super-Eddington accreting massive black holes (SEAMBHs, $\mdot\sim20$--$300$): the predicted rms declines steeply through this range, reaching $\sigma_m\simeq0.01$--$0.02$ at $\mdot\gtrsim100$. The purple points in Figure~\ref{fig:local} show the super-Eddington subsample of \citet{lu19} at $10^7$--$10^8\,M_\odot$. In the figure, we plot the data with a reliably constrained campaign, where the damping time is shorter than 10\% of the light-curve baseline. Only two sources passing this cut have SEAMBHs accretion rates; they sit at the few percent level, as predicted by the model.
(iii) Tidal disruption events (TDEs; peak $\mdot\sim10^2$--$10^3$ for $10^6$--$10^7\,M_\odot$ black holes): their optical/UV light curves are famously smooth, with residuals about the power-law decline at the level of the photometric noise, i.e., at most a few percent on day-to-month baselines \citep{gezari21}. This is consistent with our model prediction $\sigma_m\lesssim0.01$--$0.03$ in this regime. Unlike the AGN classes above, this comparison is qualitative, since TDE black holes are typically less massive than the model curves shown. (iv) Low-luminosity AGNs and LINERs ($\mdot\lesssim10^{-2}$): the band flux is split among large numbers independent zones whose inflow times greatly exceed the baseline. This gives a weak short-term variability ($\sigma_m\sim0.02$ at 1\,yr), which is recovered by 10\,yr ($\sim0.06$--$0.07$). This matches with the observed character of this class, where short-term, large-amplitude variability is rare while long-term UV variability is commonly observed \citep[][and references therein]{ho08}. Below $\mdot\sim10^{-2}$ the flow becomes RIAF \citep{yuan14}, and the model curves are truncated. For this range of accretion rate, the accretion flow geometry is no longer consistent with our current model.
(v) Local LRD analogs ($\lambda_{\rm Edd}\simeq0.1$ or $\mdot\simeq 1$): \cite{lin25} identified small sample of nearby V-shaped broad-line galaxies as local analogs of the LRDs, and subsequently, \cite{Lin26} provide their first time-domain study from $\sim$6-yr Zwicky Transient Facility(ZTF) light curves. Three sources show significant excess variance in all three bands, two are well fit by a damped random walk (structure function $SF_\infty (\equiv\sqrt{2}\sigma_{\rm d})\simeq0.18$~mag, damping timescale $\tau\simeq34$--$82$~d, consistent with the dwarf-AGN $M_{\rm BH}$--$\tau$ relation \citep{burke21}) (green stars in Figure~\ref{fig:local}). For this accretion rate regime, our model gives peak variability. These detections lie a factor of few below the model peak (the same trend-level offset as the NLS1s) and a factor of several above the super-Eddington quiet branch. The analogs thus vary like ordinary near-Eddington AGNs. In the most stringent limit, $SF_\infty<0.02$~mag for J003515+084859 is comparable to those of the high-$\tilde{z}$ LRDs despite near-Eddington accretion rate. However, \cite{Lin26} caution that this limit is most likely underestimated because of the observed cadence and short baseline of its light curve.

Finally, we would  like to mention that this comparison carries substantial caveats: the $\sigma_m$ from the model is an ensemble rms over fixed baselines, while the observations mix inhomogeneous campaign durations, DRW amplitudes, and qualitative TDE smoothness. Additionally, the normalization is anchored to the BLS1 regime, so agreement elsewhere is at the trend level. The comparison thus establishes only the sign and robustness of the high-$\mdot$ decline relevant to the LRDs.

\bibliography{sample701}{}
\bibliographystyle{aasjournalv7}

\end{document}